\documentclass[sigconf,natbib=false]{acmart}

\AtBeginDocument{%
  }

\setcopyright{acmcopyright}
\copyrightyear{2018}
\acmYear{2018}
\acmDOI{XXXXXXX.XXXXXXX}

\definecolor{brickred}{HTML}{f03b20}

\definecolor{rubinered}{HTML}{CE0058}

\RequirePackage[
  datamodel=acmdatamodel,
  style=acmnumeric,
  ]{biblatex}

\usepackage{amsmath}
\usepackage{adjustbox}
\usepackage{graphbox}
\usepackage{gensymb}
\usepackage{titlesec}

\usepackage{tikz}
\usepackage{fancyhdr}   % custom header/footer
\titlespacing{\section }{1pt}{1pt}{1pt}
\titlespacing{\subsection}{1pt}{1pt}{1pt}
\titlespacing{\subsubsection}{1pt}{1pt}{1pt}
\begin{document}

%%
%% The "title" command has an optional parameter,
%% allowing the author to define a "short title" to be used in page headers.
% \title{Indoor Localization via Bluetooth Low Energy and Inertial Measurement Unit Sensors Using a Cluster of Edge-Computing Devices in Clinical Environments}\hyeok{Maybe too long title}
\title{Indoor Localization using Bluetooth and Inertial Motion Sensors in Distributed Edge and Cloud Computing Environment}

%------------------------------------------------
% REMOVE FOOTER FOR ARXIV SUBMISSION
%------------------------------------------------
\pagestyle{fancy}
\fancyhf{}
\renewcommand{\headrulewidth}{0pt}
\AtBeginShipout{\AtBeginShipoutAddToBox{%
  \begin{tikzpicture}[remember picture, overlay, red]
    \node[anchor=south, font=\LARGE] at ([yshift=15mm]current page.south) {This manuscript is under review. Please write to yash@dbmi.emory.edu for any questions};
  \end{tikzpicture}%
}}
%------------------------------------------------

% Authors, Affiliation, and Emails in single lines.
% \numberofauthors{8}
\settopmatter{authorsperrow=4}

% \iffalse
%%%%%%%%%%%%%%%%%%%%%%%%%%%%%%%%%%%%%%%%%%%%%%%%%%%%%%%%%%%%%%%%%%%%%%%%%
%%
%% The "author" command and its associated commands are used to define
%% the authors and their affiliations.
%% Of note is the shared affiliation of the first two authors, and the
%% "authornote" and "authornotemark" commands
%% used to denote shared contribution to the research.

\author{Yashar Kiarashi}
\authornote{Equal Contribution}
\affiliation{%
  \institution{Emory University}
  % \institution{Department of Biomedical Informatics, School of Medicine, Emory University}
  % \city{Atlanta}
  % \state{Georgia}
  \country{USA}
}
\email{yash@dbmi.emory.edu}

\author{Chaitra Hedge}
\authornotemark[1]
\affiliation{%
  \institution{Georgia Tech}
  % \institution{School of Electrical and Computer Engineering, Georgia Institute of Technology}
  % \city{Atlanta}
  % \state{Georgia}
  \country{USA}
  }
\email{chegde@gatech.edu}

\author{Venkata Siva Krishna Madala}
\authornotemark[1]
\affiliation{%
  \institution{Georgia Tech}
  % \institution{School of Electrical and Computer Engineering, Georgia Institute of Technology}
  % \city{Atlanta}
  % \state{Georgia}
  \country{USA}
 }
\email{vmadala3@gatech.edu}

\author{ArjunSinh Nakum}
\affiliation{%
  \institution{Georgia Tech}
  % \institution{School of Electrical and Computer Engineering, Georgia Institute of Technology}
  % \city{Atlanta}
  % \state{Georgia}
  \country{USA}
}
\email{arjun5inh@gatech.edu}

\author{Ratan Singh}
\affiliation{%
  \institution{Georgia Tech}
  % \institution{School of Electrical and Computer Engineering, Georgia Institute of Technology}
  % \city{Atlanta}
  % \state{Georgia}
  \country{USA}
 }
\email{rsingh388@gatech.edu}

\author{Robert Tweedy}
\affiliation{%
  \institution{Emory University}
  % \institution{Department of Biomedical Informatics, School of Medicine, Emory University}
  % \city{Atlanta}
  % \state{Georgia}
  \country{USA}
}
\email{robert.tweedy@emory.edu}

\author{Gari D. Clifford}
\affiliation{%
  \institution{Emory University \& Georgia Tech}
  % \institution{Depts. of Biomed. Inform. \& Biomed. Eng., Emory University \&, Georgia Institute of Technology}
  % \city{Atlanta}
  % \state{Georgia}
  \country{USA}
  }
\email{gari@dbmi.emory.edu}

\author{Hyeokhyen Kwon}
\affiliation{%
  \institution{Emory University}
  % \institution{Department of Biomedical Informatics, School of Medicine, Emory University}
  % \city{Atlanta}
  % \state{Georgia}
  \country{USA}
}
\email{hhyeokk@emory.edu}
%%
%% By default, the full list of authors will be used in the page
%% headers. Often, this list is too long, and will overlap
%% other information printed in the page headers. This command allows
%% the author to define a more concise list
%% of authors' names for this purpose.
\renewcommand{\shortauthors}{Kiarashi et al.}
%%%%%%%%%%%%%%%%%%%%%%%%%%%%%%%%%%%%%%%%%%%%%%%%%%%%%%%%%%%%%%%%%%%%%%%%%
% \fi

%%
%% The abstract is a short summary of the work to be presented in the
%% article.

\begin{abstract}

Spatial navigation of indoor space usage patterns reveals important cues about the cognitive health of individuals.
In this work, we present a low-cost, scalable, open-source edge computing system using Bluetooth Low Energy (BLE) and Inertial Measurement Unit sensors (IMU) for tracking indoor movements for a large indoor facility (over 1600 m$^2$) that was designed to facilitate therapeutic activities for individuals with Mild Cognitive Impairment.
The facility is instrumented with 39 edge computing systems with an on-premise fog server, and subjects carry BLE beacon and IMU sensors on-body.
% The proposed system has 39 edge computing systems distributed across the study site placed in the ceiling that is connected to an on-premise computing and data storage server through private networks.
% While subjects carry BLE and IMU sensors on the weist, the data is streamed to surrounding edge computing devices, which data is transferred to on-premise server for localization of individuals.
We proposed an adaptive trilateration approach that considers the temporal density of hits from the BLE beacon to surrounding edge devices to handle inconsistent coverage of edge devices in large spaces with varying signal strength that leads to intermittent detection of beacons.
%unstable Received Signal Strength Indicator (RSSI) signals.
The proposed BLE-based localization is further enhanced by fusing with an IMU-based tracking method using a dead-reckoning technique. 
Our experiment results, achieved in a real clinical environment, suggest that an ordinary medical facility can be transformed into a smart space that enables automatic assessment of the individual patients' movements. 

\end{abstract}

% 228 words

%%
%% The code below is generated by the tool at http://dl.acm.org/ccs.cfm.
%% Please copy and paste the code instead of the example below.
%%

% \ccsdesc[500]{Computer systems organization~Embedded systems}
% \ccsdesc[300]{Computer systems organization~Redundancy}
% \ccsdesc{Computer systems organization~Robotics}
% \ccsdesc[100]{Networks~Network reliability}

%%
%% Keywords. The author(s) should pick words that accurately describe
%% the work being presented. Separate the keywords with commas.

% \received{20 February 2007}
% \received[revised]{12 March 2009}
% \received[accepted]{5 June 2009}

%%
%% This command processes the author and affiliation and title
%% information and builds the first part of the formatted document.
\maketitle

\section{Introduction}

Over the past decade, the presence of smart connected devices has led to a new wave of ambient monitoring of patients in clinical environments. 
Patients' movements in indoor spaces play a vital role in assessing health, particularly cognitive health~\cite{ghosh2022machine}. 
A range of solutions have been proposed to support tracking a subject's location using various sensors: radio-frequency identification (RFID)~\cite{vuong2014automated}, infrared (IR)~\cite{cheol2017using}, WiFi~\cite{van2016performance}, and Bluetooth~\cite{yoo2018real}.
Among them, BLE-based solutions gained popularity for their unique advantages, being low-cost, low-power, and privacy-preserving solutions.
% Over the past decade, the presence of smart connected devices has led to a new wave of ambient monitoring of patients in clinical environments. 
% Patients' movements in indoor spaces plays a vital role in assessing health, particularly cognitive health~\cite{ghosh2022machine}. 
% A range of solutions have been proposed to support tracking subject whereabouts using various sensors: radio-frequency identification (RFID)~\cite{vuong2014automated}, infrared (IR)~\cite{cheol2017using}, WiFi~\cite{van2016performance}, and Bluetooth~\cite{yoo2018real}.
% Among them, BLE-based solution gained popularity for its unique advantages being low-cost, low-power, and privacy-preserving solutions.
% Due to the low cost and scalability of Bluetooth Low Energy (BLE) transmitters, or beacons, numerous work proposed using BLE-based subject localization and tracking in large indoor environments.

Although with many successes in  BLE-based localization systems, it is yet to be validated for large indoor spaces (1600 m$^2$), which challenges the uniform distribution of BLE receivers~\cite{bai2020low,qureshi2019evaluating}.
Conventionally, the Received Signal Strength Indicator (RSSI) is used for BLE-based localization, representing distances between BLE beacons/transmitters~\cite{newman2014apple}.
In wide indoor spaces having multiple rooms and furniture with varying materials, RSSI can be inconsistent and unstable across the area due to environmental factors, including absorption, diffraction, reflection, interference, etc., which significantly hampers the localization accuracy~\cite{townsend2014getting}.
% Although with many successes in  BLE-based localization systems, it is yet to be validated for large indoor spaces (18000 sqft), which challenges the uniform distribution of BLE receivers across the space~\cite{bai2020low,qureshi2019evaluating}.
% Conventionally, the Received Signal Strength Indicator (RSSI) is used for BLE-based localization, which represents distances between a BLE beacons/transmitters~\cite{newman2014apple}.
% In wide indoor spaces having multiple rooms and furniture with varying materials, RSSI signal can be inconsistent and unstable across the space due to environmental factors, including absorption, diffraction, reflection, interference, etc, which significantly hampers the localization accuracy~\cite{townsend2014getting}.

In this study, we validate the feasibility of using a BLE- and IMU-based indoor localization system in an indoor space of over 1600 m$^2$,
% 18,000 square feet. 
designed to carry out therapeutic services for individuals with Mild Cognitive Environment (MCI).
MCI is a clinically-recognized stage between normal aging and dementia, marked by a decline in cognitive functions like memory and attention. 
The environment aimed to encourage MCI patients to participate in various therapeutic activities, including exercise and memory training, while fostering social interaction opportunities (e.g., during meals and in lounge areas). 
Automated patient localization and tracking system can provide important clues about their cognitive health while performing a wide range of activities.

%To track movements in a wide space with a complex structure, we designed an edge computing framework that comprises 39 edge computing units, each containing a Raspberry Pi 4 Model B (4GB of RAM) equipped with a BLE 4 antenna that is placed on the ceiling, connected via a private wired network to an on-premise fog server. 
%Each participant wore a belt bag containing a BLE beacon (Smart Beacon SB18-3 Kontact.io) and a nine-axis IMU (the Yost 3-Space Data Logger).
%While patients walk, RSSI signals of the BLE beacon are captured by the edge devices within range, which are processed in real-time to localize the subject.
% In this study, we validate the feasibility of using BLE-based indoor localization system for the indoor space spanning 18,000 square feet.
% Our study site is designed to carry out theraputic services for individuals with Mild Cognitive Environment (MCI), that consists of various functioning spaces including gym, kitchen, library, and so on.
% To track movements in a wide space with complex structure, we designed an edge computing framework that comprises 39 edge computing units, each containing a Raspberry Pi 4 Model B (4GB of RAM) equipped with a BLE 4 antenna that is placed on the ceiling, connected via a private wired network to an on-premise fog server. 
% Each participant wore a belt bag containing a BLE beacon (Smart Beacon SB18-3 Kontact.io) and a nine-axis IMU (the Yost 3-Space Data Logger).
% While patients walk, RSSI signals of the BLE beacon is captured by the  edge devices within range, which is processed real time to localize the subject.

\begin{figure*}[t]
    \centering
    \begin{adjustbox}{width=0.9\linewidth,center}
    \begin{tabular}{c c c}
         \includegraphics[align=c, height=0.2\textwidth]{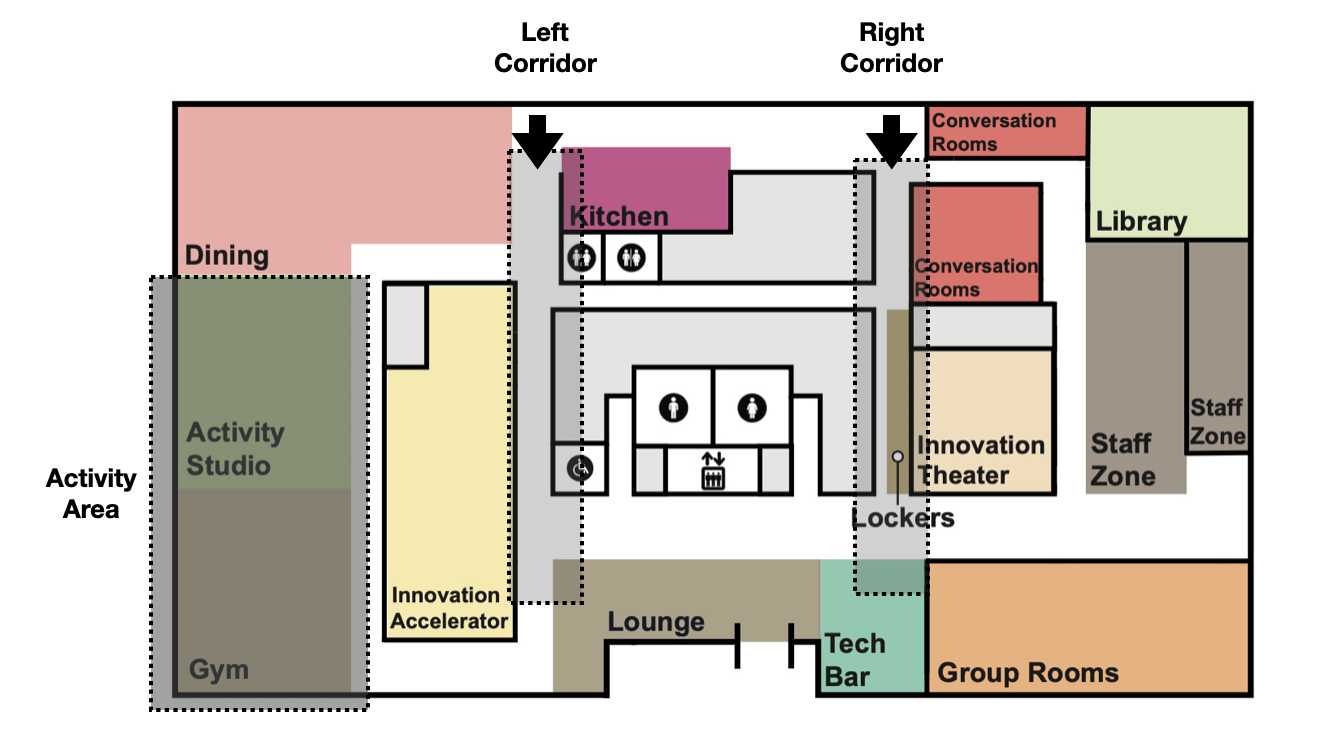} &
         \includegraphics[align=c, height=0.16\textwidth]{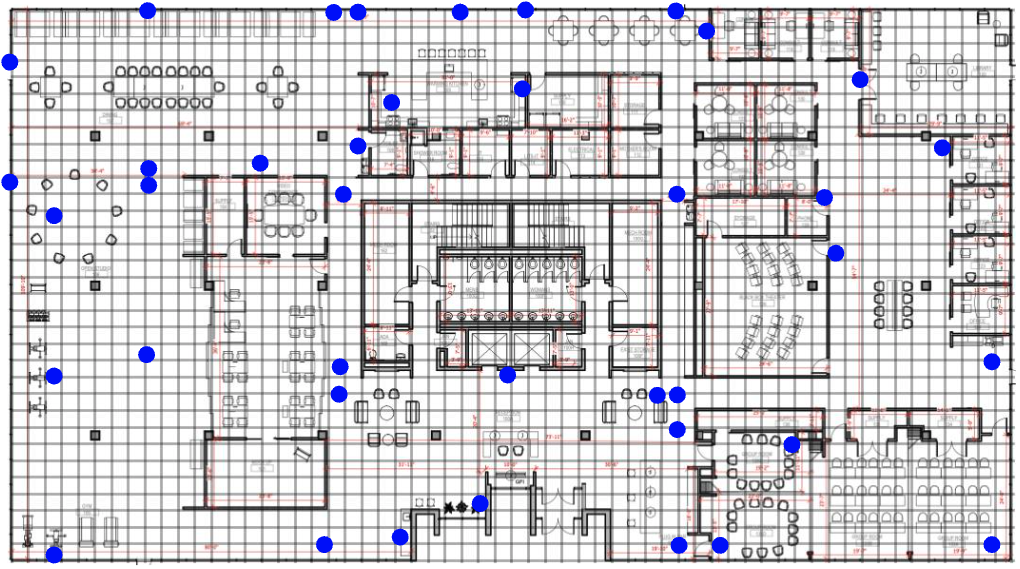} &
         \includegraphics[align=c, height=0.16\textwidth]{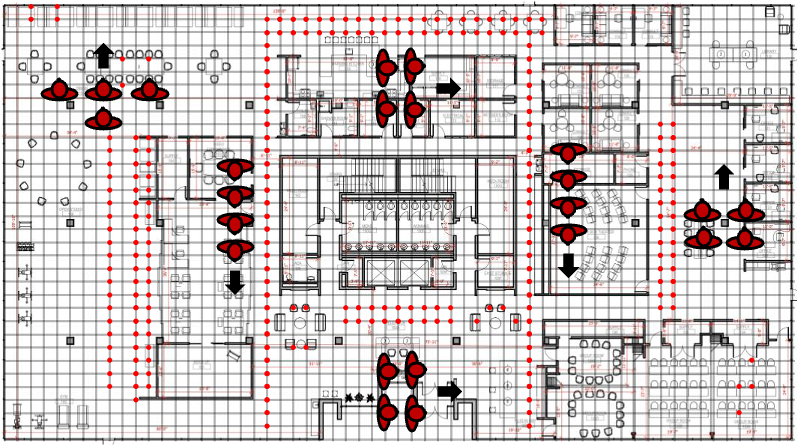} \\
         (a) & (b) & (c)
    \end{tabular}   
    \end{adjustbox}
    \vspace{-0.13in}
    \caption{Study site and data collection.
    (a) Study sites are designed to include various utility spaces.
    (b) The locations of 39 edge computing devices (Raspberry Pi v4 model B) in the ceiling.
    (c) The location of subjects in our data collection.}
    \vspace*{-.0in}
    \label{fig:study_site}
\end{figure*}

\iffalse
\begin{figure}[t]
\centering
\includegraphics[align=c, width=1.0\linewidth]{figures/study_site.png}
\caption{ Study site space utilization. The space partitioned into gym, activity studio, dining area, innovation accelerator, lounge, kitchen, tech bar, lockers, group rooms, innovation theater, conversation rooms library, staff zone, and other transition zones.}
\label{fig:study_site}
\centering
\end{figure}
\fi

The proposed BLE-based localization method, namely adaptive trilateration, uses temporal information of RSSI, which is the frequency of signals received from the beacons over time, that could effectively tackle the fluctuations in RSSI due to the environmental factors abovementioned.
Additionally, integrating with IMU-based dead reckoning, the proposed approach reliably and accurately localized multiple subjects continuously over time.
The proposed privacy-preserving framework is also cost-effective and easy to install owing to low-cost edge computing devices (approximately \$200 per device).
Our system is versatile and can be applied to analyze the movement and interactions of individuals in various contexts, such as healthcare facilities, where this study was conducted,
We expect the proposed framework to transform typical indoor spaces into smart environments, providing low-cost, opportunistic services for ambient behavior assessments.

\section{Related Work}

% \subsection{Indoor localization with BLE}

\textit{Indoor localization with BLE:}
% Localization using BLE RSSI is an active research area for indoor application, as Global Positioning System (GPS) requires a line-of-sight connection between the transmitter and receiver to ensure acceptable accuracy~\cite{riady2022indoor,basri2016survey,atashi2021orientation,ramadhan2020applying}.
Indoor Localization using BLE is an active research area and is deployed largely in two different approaches:
% ~\cite{riady2022indoor,basri2016survey,atashi2021orientation,ramadhan2020applying}.
% RSSI-based localization techniques are deployed largely in two different approaches: 
% proximity, 
fingerprinting 
and trilateration~\cite{spachos2018microlocation,zafari2015microlocation,riady2022indoor,basri2016survey,atashi2021orientation,ramadhan2020applying}.
% Proximity approach is limited in localizing the absolute location of the subject, and only distinguishes the presence and the distance of the subject in the signal range.
Fingerprinting can precisely localize subjects in absolute locations, which is widespread for its simplicity and bounded positioning error.
However, it requires a detailed reference radio grid map assuming that BLE beacons are in fixed spaces.
Otherwise, the radio grid map needs to be recalibrated whenever the BLE beacons are moved~\cite{chai2007reducing,guzman2015integration}. In our study site, our edge devices had to be replaced or repositioned frequently so fingerprinting was not applicable.

Trilateration can localize subjects' relative location based on known locations of multiple (3+) BLE beacons using triangulation.
Although conceptually simple, the major challenge is that it uses the path loss model to estimate the subject's distance (BLE scanner) from each BLE beacon based on signal strengths.
As a result, RSSI needs to be stable to estimate an accurate and consistent location of subjects.
Yet, RSSI can fluctuate due to environmental factors such as absorption, diffraction, or interference, which severely degrades localization accuracy.
Specifically, in our study, edge devices were placed/hidden in the ceiling where complex metallic structures were present, which made the signal fluctuate, creating inconsistent measurements according to the signal directions.
To tackle this challenge, we proposed an adaptive trilateration method that uses the temporal context of RSSI while subjects walk.
% Our study adapted trilateration, as we only needed to keep where each edge device is for identifying absolute location of subjects in the study space.

% \subsection{Sensor Fusion for Indoor Localization}

\textit{Sensor Fusion for Indoor Localization:}
Many works proposed using other wireless or on-body sensors along with BLE beacons~\cite{kolakowski2017kalman,gang2019smartphone} to improve the localization accuracy.
WiFi is widely used with BLE as it is easily available in any indoor environment nowadays~\cite{antevski2016hybrid,kanaris2017fusing}.
Previously, WiFi and BLE hybrid systems were used to optimize fingerprinting with fuzzy logic models~\cite{al2019fuzzy}.
% WiFi and BLE hybrid systems were used to optimize the placement of WiFi access points and BLE beacons~\cite{tian2020optimizing}, or to improve fingerprinting with fuzzy logic models\cite{al2019fuzzy}.
% Tian \etal~\cite{tian2020optimizing} proposed a hybrid localization approach using WiFi and BLE signals that can optimize the placement of WiFi access points and BLE beacons that significantly impacts localization performance.
% Al-Madani \etal~\cite{al2019fuzzy} proposed fingerprinting-based localization method with fuzzy logic model by fusing WiFi and BLE sensors.
Although effective, we did not use WiFi signals as it was disabled in our edge devices for security reasons.
For on-body sensors, IMU sensors are widely used for inertial dead reckoning, which estimated heading and step counts for deriving  trajectory~\cite{huang2019hybrid,jamil2021enhanced,meliones2018blind,chen2012infrastructure}.
% More recently, a number of work proposed machine learning-based BLE and IMU fusion method that is based on handcrafted features and shallow models~\cite{atashi2021orientation} or multi-modal encoder-decoder neural network models~\cite{hua2022smartfps}. 
% Atashi~\etal~\cite{atashi2021orientation} proposed a classification model that can estimate the every 45\degree orientation of BLE device fusing RSSI signals from multiple beacons and IMU sensors.
% Hua and Yang~\etal~\cite{hua2022smartfps} proposed a multi-modal encoder-decoder network to fuse wireless and inertial movements.
In this work, we use the standard dead reckoning approach for IMU sensor on the subject's waist that is fused with BLE-based localization to improve localization accuracy~\cite{woodman2008pedestrian,robesaat2017improved}.

\section{Indoor Localization fusing BLE and IMU}

\textit{Study Environment:}
The therapeutic space consists of various functioning spaces, including a gym, kitchen, library, dining area, cinema, maker lab, and other bespoke rooms/spaces designed to improve cognition for individuals with MCI. See \autoref{fig:study_site} (a). 
To track movements in a wide space with a complex structure, we designed an edge computing framework that comprises 39 edge computing units, each containing a Raspberry Pi 4 Model B (4GB of RAM) equipped with a BLE 4 antenna that is placed on the ceiling~\cite{s22072511,kwon2023indoor}, shown in \autoref{fig:study_site} (b), where network and power sources are accessible nearby.
% , connected via a private wired network to an on-premise, Health Insurance Portability and Accountability Act (HIPAA)-compliant fog server\cite{s22072511}. 
While at the study site, each participant wore a belt bag containing a BLE beacon (Smart Beacon SB18-3 Kontact.io) and a nine-axis IMU (the Yost 3-Space Data Logger).
As subjects walk around, edge computing devices in the range read RSSI from BLE Beacon and accelerometer, gyroscope, and magnetometer from IMU sensors, which data is transmitted to an on-premise, Health Insurance Portability and Accountability Act (HIPAA)-compliant fog server connected via a private wired network that runs localization algorithms, as shown in \autoref{fig:Flowchart}. 
We employed the Bleak \cite{Bleak} python package and scanned BLE beacons with the sampling frequency of $2 Hz$.
% The data was collected from the multiple edge computing nodes over the day and transmitted to an on-premise, Health Insurance Portability and Accountability Act (HIPAA)-compliant fog computing node
% , which runs localization algorithms, as shown in \autoref{fig:Flowchart}.
This centralized approach to network storage and localization allows for efficient and accurate management of edge devices, making it an essential component of any privacy-preserving system.

% \subsubsection{BLE-based Localization}
\textit{BLE-based Localization:}
The RSSI value of a BLE signal represents the strength of the signal transmitted by the BLE Beacon and can estimate the distance between the transmitter and receiver.
Our beacon has a broadcasting power of -12 dBm.
The distance between a beacon and edge device in meters, $d$, is calculated from the measured RSSI value as follows:
% The RSSI value of a BLE signal represents the strength of the signal transmitted by the BLE Beacon and can provide an estimate of the distance between the transmitter and receiver.
% Our beacon has broadcasting power of -12 dBm.
% The distance between a beacon and edge device in meters, $d$, is calculated from the measured RSSI value:

% \vspace*{-0.2in}

\begin{equation}
    \scalebox{1.2}{$d =\ 10^\frac{ M_{RSSI} - I_{RSSI}}{10N}$}
    \label{eq:1}
% \vspace*{-0.1in}
\end{equation}

\noindent Where $M_{RSSI}$ (i.e., 1m RSSI) is the RSSI at a distance of 1 m from the antenna, which the manufacturer gives, and $I_{RSSI}$ is the instantaneous RSSI measured by the receiver in Decibels (dB).
$N$ represents environmental surroundings near and between the receiver and transmitter, which is a function of the medium for electromagnetic wave propagation, such as air ducts or metal structures in the ceiling where edge computing devices are located.
We manually derived $N$ by measuring $M_{RSSI}$ for each edge computing device, as different metallic structures surrounded each device in the ceiling.
% \noindent where $MRSSI$ (i.e., 1m RSSI) be the received RSSI at a distance of 1 meter from the antenna, which is given by the manufacturer, and $Instant RSSI$ is RSSI measured by the receiver in Decibel (dB).
% $N$ represents environmental content, which is a function of the medium for electromagnetic wave propagation.
% We manually derived $N$ by measuring $MRSSI$ for each edge computing device, as each device was surrounded by different metallic structures in the ceiling. %\hyeok{How did you fine-tuned $N$?}

Next, the derived distance from the beacon to each edge computing node, $d$, is integrated to determine subjects' locations adapting a trilateration-based approach~\cite{sadowski2018rssi}.
% , a popular BLE-based localization approach.
% Trilateration uses the principle of triangulation, which uses the distance from three+ known points to determine the location of an object.
In this work, we propose an adaptive trilateration that additionally considers temporal information, which was found to be robust in noisy environments with sparse BLE receiver distributions.
The signals fluctuated due to the metallic objects inside the ceilings, where edge computing devices were located. 
This severely degraded the performance of the standard trilateration, which heavily relies on accurate distance measurement for points. 

% Next, the distance from the beacon to each edge computing node is integrated to determine subjects' locations adapting trilateration-based approach~\cite{sadowski2018rssi}, a popular BLE-based localization approach.
% Trilateration uses the principle of triangulation that uses the distance from three+ known points to determine the location of an object.
% In this work, we propose an adaptive trilateration that additionally considers temporal RSSI information, which we empirically found to be robust in noisy environments like ours.
% The RSSI signals fluctuated due to the metallic objects inside the ceilings, where edge computing devices were located. 
% This severely degraded the performance of the standard trilateration, which heavily relies on accurate distance measurement for points. 

\begin{figure}[t]
    \centering
    \includegraphics[align=c, width=0.6\linewidth]{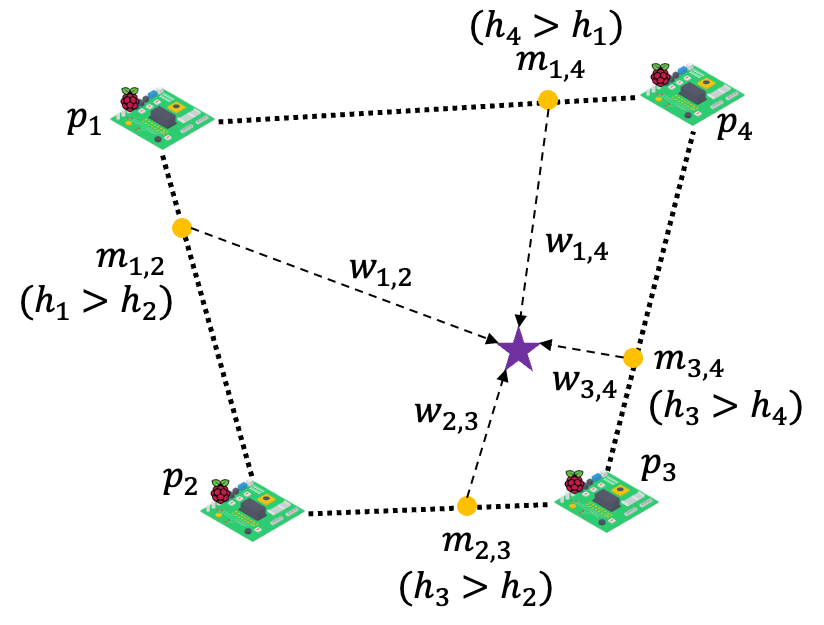}
    \vspace{-0.2in}
    \caption{Adaptive trilateration first calculates potential locations, $m_{i,j}$, between each Pis, $p_i$ and $p_j$, based on the hits, $h_i$ and $h_j$, where $i,j={1, 2, 3, 4}$.
    Then, the $m_{i,j}$s are aggregated with weights, $w_{i,j}=h_i+h_j$, to localize the subject (\textcolor{purple}{star}).}
    \vspace{-0.0in}
    \label{fig:adapt_tri}
\end{figure}

Shown in \autoref{fig:adapt_tri}, the proposed method considers the weighted average of the distances of nearby edge devices from the subject's beacon within a certain time frame.
For every $t$ time window, we calculate the Received Hits Strength Index (RHSI) for every edge device in the range of a beacon, $p_1, p_2, \cdots, p_n$, which counts the number of hits for each edge device, $h_1, h_2, \cdots, h_n$. 
We empirically found $t=10sec$ sufficient to consider meaningful numbers of hits with sufficiently strong RSSI values across $p_i$s.
Due to the fluctuations of RSSI from metallic structures in the ceiling, the hit made from beacons to each edge device was inconsistent, although with $2Hz$ sampling protocols.
% Also, we only considered RSSI hits above the average value of RSSI observed at each edge device, where threshold values were manually derived through data collection around each device.

% Shown in \autoref{fig:adapt_tri}, the proposed method considers the weighted average of the distances of nearby edge devices from the beacon within a time window.
% For every $t$ time window, we calculate Received Hits Strength Index (RHSI) for every edge device in the range of a beacon, $p_1, p_2, \cdots, p_n$, which counts the number of RSSI hits for each edge device, $h_1, h_2, \cdots, h_n$. 
% We empirically found $t=30sec$ to be sufficient to consider meaningful numbers of hits with sufficiently strong RSSI values across $p_i$s.
% Due to the fluctuations of RSSI coming from metallic structures in the ceiling, the hit made from beacons to each edge device was not consistent, although with $2Hz$ sampling protocols.
% Also, we only considered RSSI hits above the average value of RSSI observed at each edge device, which was manually derived through data collection around each device.

\begin{figure}[t]
\centering
\includegraphics[align=c, width=0.9\linewidth]{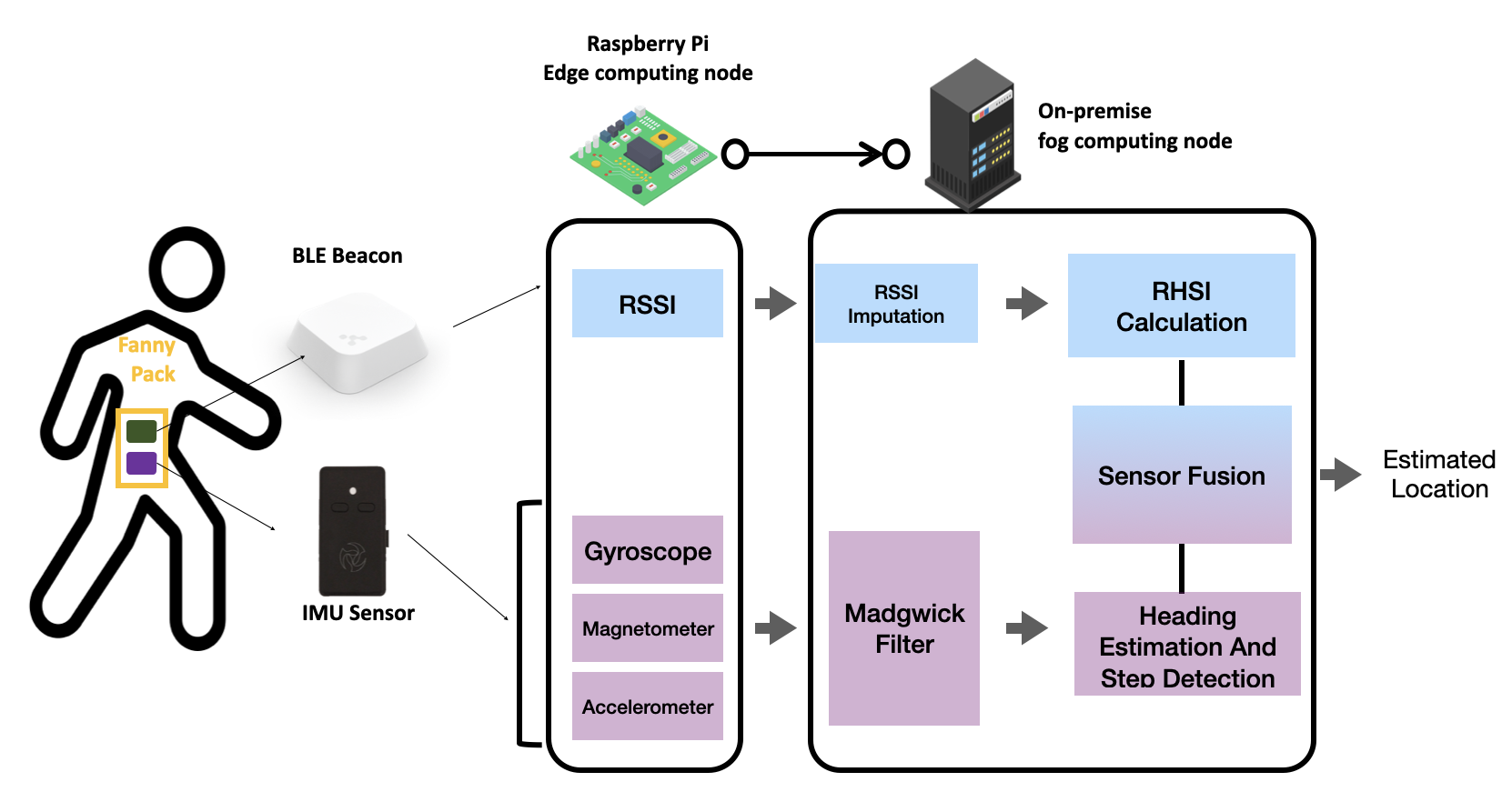}
\vspace{-0.2in}
\caption{Overall diagram for the proposed indoor localization system using BLE and IMU fusion.} 
% \vspace*{-.15in}
\label{fig:Flowchart}
\end{figure}

\iffalse
\begin{figure}[H]
\centering
\includegraphics[scale=0.3]{figures/Sensor_fusion_felowchart.png}
\caption{Flowchart illustrating the integration of IMU and BLE technologies for enhanced localization accuracy.} 
\label{fig:Flowchart}
\end{figure}
\fi

Next, we integrate all RSSI values from all edge devices to localize subjects.
First, we derived the potential locations of each subject considering pairs of edge devices.
For example, the estimated location between $p_i$ and $p_j$, which had $h_i$ and $h_j$ hits for the past $10 sec$, is $m_{i,j}=\frac{h_ip_i+h_jp_j}{h_i+h_j}$.
We find this pair-wise location for all pairs of edge devices within the beacon range.
Once we find $m_{1,2}, m_{1,3}, m_{1,4}, \dots, m_{n-1,n}$ for all edge device pairs, we calculate the overall weighted centroid using the total number of hits observed in a pair of the device combined, $w_{i,j} = h_i + h_j$.
The final estimated location based on the BLE signal is $L_{BLE} = \frac{\sum^n_i\sum^n_j w_{i,j}m_{i,j}}{\sum^n_i\sum^n_j w_{i,j}}$.

% Next, we integrate all RSSI hit information from all edge devices to localize subjects.
% First, we derived potential locations of each subject considering pairs of edge devices.
% For example, the estimated location between $p_i$ and $p_j$, which had $h_i$ and $h_j$ hits for the past $30 sec$, is $m_{i,j}=\frac{h_ip_i+h_jp_j}{h_i+h_j}$.
% We find this pair-wise location for all pairs of edge device within the range of beacon.
% Once we find $m_{1,2}, m_{1,3}, m_{1,4}, \dots, m_{n-1,n}$ for all edge device pairs, we calculate the overall weighted centroid using the total number of hits observed in a pair of the device combined, $w_{i,j} = h_i + h_j$.
% The final estimated location based on BLE signal is $L_{BLE} = \frac{\sum^n_i\sum^n_j w_{i,j}m_{i,j}}{\sum^n_i\sum^n_j w_{i,j}}$.

% \subsubsection{BLE and IMU Fusion}
\textit{BLE and IMU Fusion:}
Next, we fuse BLE-based localization with IMU-based tracking methods to improve the tracking and localization quality.
First, we applied a standard dead-reckoning technique for an IMU sensor on the subjects' weist~\cite{liu2020novel,salimibeni2021iot}. 
We first denoise the signals by applying a low-pass filter and then estimate step counts and lengths from zero-crossings in accelerometry signals.
The headings are estimated using Madgwick filter~\cite{madgwick2011estimation} from accelerometers, gyroscopes, and magnetometers.

The computed IMU-based trajectory is integrated with BLE-based localization with equal weight on each modality.
Given a location estimated at time $t-1$ using both BLE and IMU, $[x^{t-1}_{BLE+IMU}$, $y^{t-1}_{BLE+IMU}]$, we estimate the potential location of a subject when only considering the IMU trajectory between time $t-1$ and $t$, which is $[x^t_{IMU}, y^t_{IMU}]=[x^{t-1,t}_{IMU}+x^{t-1}_{BLE+IMU}, y^{t-1,t}_{IMU}+y^{t-1}_{BLE+IMU}]$.
Once we have BLE-based location at time $t$, $[x^{t}_{BLE}, y^{t}_{BLE}]$, then we take average with IMU-based estimated location at time $t$, where BLE and IMU fused location is $[x^{t}_{BLE+IMU}, y^{t}_{BLE+IMU}]=[\frac{x^t_{IMU}+x^{t}_{BLE}}{2}$, $\frac{y^t_{IMU}+y^{t}_{BLE}}{2}]$.
In summary, the proposed sensor fusion-based method complements BLE's sporadic ($2 Hz$) but relatively smooth localization considering $10sec$ window with IMU's dense ($42 Hz$) yet noisy trajectory estimation.
% from the scikit-kinematic library\cite{skinematics}.

% the script we used in named as imutracker.py and located in the wearable tracking repo which was migrated to the clifford lab

% Kalman filter integrated into firmware provided by the manufacturer. 
% \hyeok{Yash, do you have equations for heading calculation? Maybe Yost do not release it?}\yash{we wrote a script using  Scikit-kinematic library and didnt use the built it heading estimation. I revised the texts and added references accordingly.}

% \yash{Hyeok, what does 30s refer to?}\hyeok{This is for the window size for beacon hits, which I considered it similar to smoothing operation.}

% \hyeok{Yash, is this correct? This is how I understood your text. This does not look like a complementary filter. Could you provide equation, please, in case I am misunderstanding anything?}\yash{Yes, correct. One common approach is to use a weighted average where the weight is determined by the confidence in each sensor's measurement. But I agree that the complementary filter might not be the best term we can use. I updated the figure 2 acoordingly.}

\section{Experiment and Results}

\begin{figure*}
    \centering
    \begin{adjustbox}{width=0.62\linewidth,center}
    \begin{tabular}{c c}
        \includegraphics[align=c, width=0.4\linewidth]{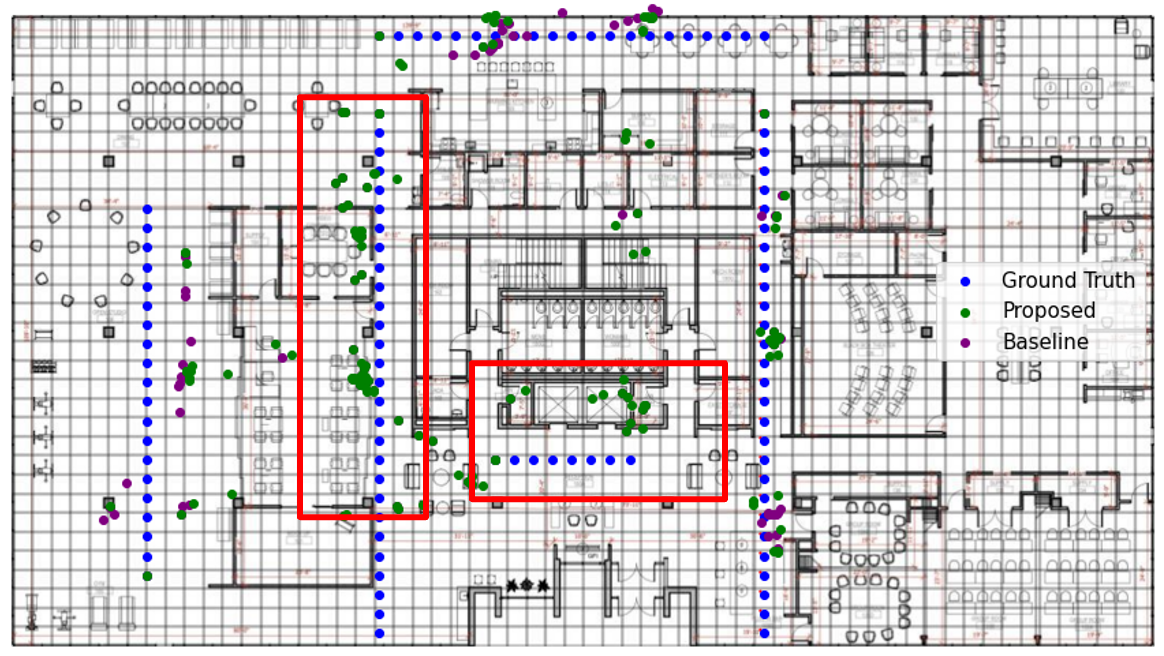} &
        \includegraphics[align=c, width=0.4\linewidth]{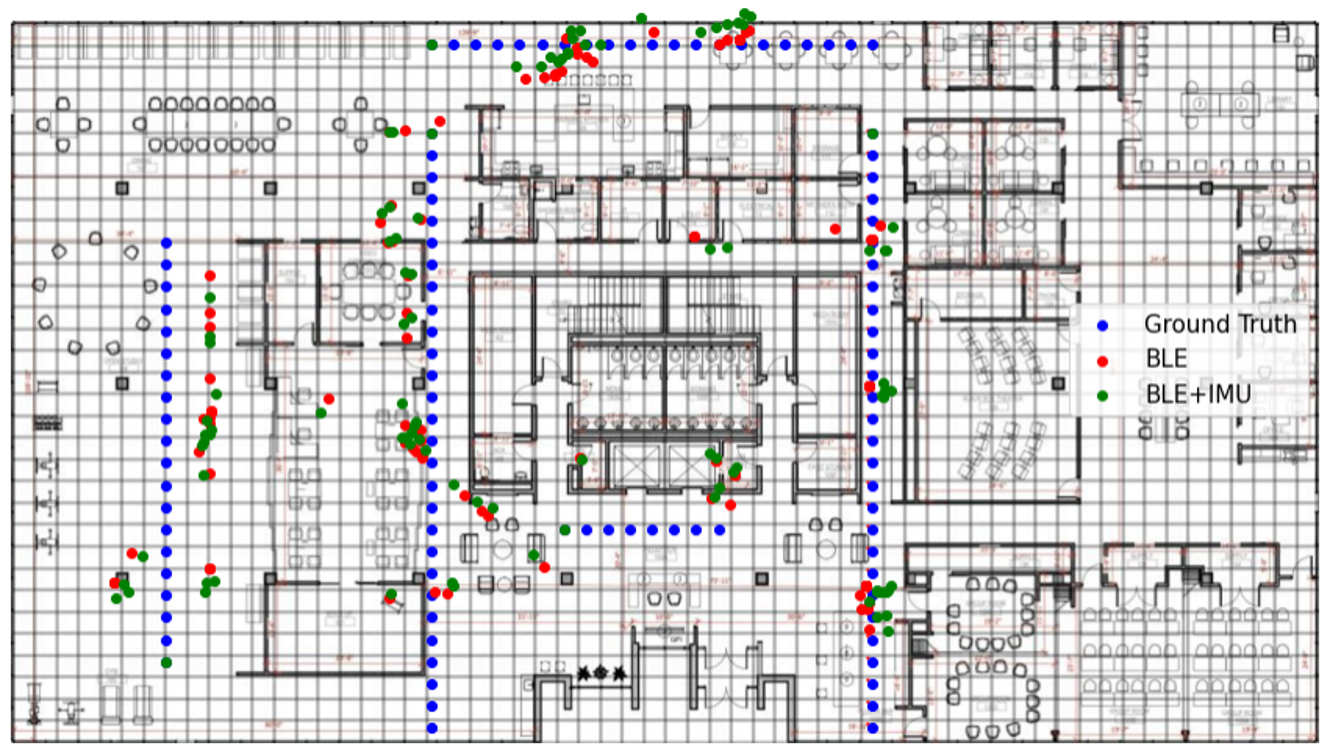} \\
        (a) & (b)
    \end{tabular}
    \end{adjustbox}
    \vspace{-0.15in}
    \caption{Qualitative results while tracking a subject in our study site.
    (a) BLE-based localization from baseline method (\textcolor{purple}{Purple}) and the proposed method (\textcolor{green}{Green}).
    % Localization results from the baseline method are severely affected by missing detections as it only considers instantaneous BLE beacon hits for trilateration (\textcolor{red}{Red} box).
    (b) Localization results from the proposed method when only using BLE (\textcolor{red}{Red}) and BLE and IMU fusion (\textcolor{green}{Green}).
    % As shown in the quantitative results, IMU fusion demonstrated marginally similar localization performance with BLE-only localization results.    
    Ground-truth locations are shown in (\textcolor{blue}{Blue}) for both figures.
    }
    % \vspace*{-.2in}
    \label{fig:Qualitative}
\end{figure*}

% \hyeok{Could you add details of the baseline method (standard Trilateration), please? Any citation you are specifically referring to as well?}\yash{I revised the text and added the reference for Trilateration}
% \textit{Benchmark Data Collection:}
Our benchmark dataset involved three to four individuals walking and interacting as instructed in the study site, shown in \autoref{fig:study_site} (c). 
To ensure accurate annotation of the subjects' locations and corresponding timestamps, we placed 1-meter markers throughout the entire space. An observer manually labeled the closest marker location of each subject while they walked. 
With the benchmark datasets, we evaluate the proposed method for five main regions within the study site shown in \autoref{fig:study_site} (a): Right Corridor, Left Corridor, Kitchen, Lounge, and Activity Area. 

We evaluated the performances on two levels: positioning error and room-level localization.
The positioning error was calculated as Euclidean distance in meters between the estimated and ground-truth locations to understand the error margin of tracking fine-grained movements.
% for $10s$ time frames.
% We evaluated the performances in two levels: positioning error and room-level localization.
% The positioning error was calculated as euclidean distance in meters between the estimated and ground-truth locations to understand the error margin of tracking fine-grained movements.
The room-level localization was computed as the percentage of times each area/room was predicted correctly while the subject moved within the area.
As much as granular movements are essential, understanding room-dwelling profiles can help understand the routine activities of individuals.
% The room-level localization was computed as the percentage of times each area/room was predicted correctly while subject move within the area.
% As much as granular movements are important, understanding room-dwelling profiles can help understand the routine activities of individuals.
% Overall, our evaluation protocol will demonstrate the potential granularity of movement analysi of subjects while they stay in the study site.
As a baseline method, we used a standard trilateration approach for BLE-based localization~\cite{wang2013bluetooth} that only uses concurrent beacon hits available at a time for comparison with our proposed method.

\begin{table}[t]
    \centering
    \Huge
    \caption{Positioning error (in meters) and room level localization accuracy (in \%) in different areas using BLE or both BLE and IMU.
    % \hyeok{Please include results with standard trilateration}
    % \yash{I included the standard trilateration results} 
    }
    \vspace*{-0.1in}
    \begin{adjustbox}{width=\columnwidth,center}
    \begin{tabular}{c||c c c c c||c}
         & \textbf{Right}	& \textbf{Left} &  &  & \textbf{Activity} & \\ 
         \textbf{Sensor} & \textbf{Corridor}	& \textbf{Corridor} & \textbf{Kitchen} & \textbf{Lounge} & \textbf{Area} & \textbf {Average}\\ 
         \hline\hline
         \multicolumn{7}{c}{Standard Trilateration}\\
         \hline\hline
         \multicolumn{7}{c}{Positioning error (m)}\\
         \hline\hline        
         BLE only & 5.19 & 4.82 & 4.66 & 6.32  & 6.32 & 5.97 \\
        BLE \& IMU & 4.82 & 4.74 & 4.3 & 6.53 & 5.53 & 5.46\\
        \hline\hline
        \multicolumn{7}{c}{Room Level Accuracy (\%)}\\
        \hline\hline
        BLE only & 96 & 68.6 & 73 & 50  & 81.7 & 73.8\\
        BLE \& IMU & 97.3 & 81 & 82.6 & 60.2  & 88.3 & 82.2\\
         \hline\hline
         \multicolumn{7}{c}{Adaptive Trilateration}\\
         \hline\hline
         \multicolumn{7}{c}{Positioning error (m)}\\
         \hline\hline        
         BLE only &4.33 &3.64 &3.68 &4.22  &4.64 &4.22 \\
        BLE \& IMU &4.01 &3.55 &4.4 &3.63 &3.63 &3.9\\
        \hline\hline
        \multicolumn{7}{c}{Room Level Accuracy (\%)}\\
        \hline\hline
        BLE only &100 &76.3 &77.6 &58.3  &95 &81.4\\
        BLE \& IMU &100 &84.6 &83.6 & 70.6 &96.7 &87.1\\
        \hline                 
    \end{tabular}
    % \begin{tabular}{c||c c c c c||c}
    %      & \textbf{Right}	& \textbf{Left} &  &  & \textbf{Activity} & \\ 
    %      \textbf{Sensor} & \textbf{Corridor}	& \textbf{Corridor} & \textbf{Kitchen} & \textbf{Lounge} & \textbf{Area} & \textbf {Average}\\ 
    %      \hline\hline
    %      \multicolumn{7}{c}{Standard Trilateration}\\
    %      \hline\hline
    %      \multicolumn{7}{c}{Positioning error (m)}\\
    %      \hline\hline        
    %      BLE only & 6.15 & 7.50 & 5.27 & 4.06  & 8.25 & 6.25 \\
    %     BLE \& IMU & 5.23 & 6.78 & 5.42 & 3.61 & 5.92 & 5.39\\
    %     \hline\hline
    %     \multicolumn{7}{c}{Room Level Accuracy (\%)}\\
    %     \hline\hline
    %     BLE only & 85.2 & 83.7 & 89.1 & 89.5  & 76.4 & 84.8\\
    %     BLE \& IMU & 87.3 & 84 & 85.4 & 91.2  & 88.2 & 87.2\\
    %      \hline\hline
    %      \multicolumn{7}{c}{Adaptive Trilateration}\\
    %      \hline\hline
    %      \multicolumn{7}{c}{Positioning error (m)}\\
    %      \hline\hline        
    %      BLE only &5.01 &2.94 &3.13 &4.68  &4.11 &3.97 \\
    %     BLE \& IMU &4.31 &3.43 &2.57 &4.71 &3.21 &3.65\\
    %     \hline\hline
    %     \multicolumn{7}{c}{Room Level Accuracy (\%)}\\
    %     \hline\hline
    %     BLE only &88.1 &91.4 &92.7 &89.2  &90.2 &90.3\\
    %     BLE \& IMU &90.7 &90.6 &93 & 90.4  &91.6 &91.2\\
    %     \hline                 
    % \end{tabular}
    \end{adjustbox}
    \vspace*{-0.1in}
    \label{tab:PositioningErrorOverWindowSizes}
\end{table}

\iffalse
\begin{table}[t]
\caption{Positioning error for three participants in different areas of the facility using BLE and IMU. 
\label{table:PositioningErrorOverWindowSizes}}
\resizebox{\textwidth}{!}{% 50% of the text width
\begin{tabular}{cccccccl}
\toprule
\textbf{Signal Modality}&\textbf{Right Corridor}	& \textbf{Left Corridor} & \textbf{Kitchen} & \textbf{Lounge} & \textbf{Activity Area}& \textbf {Average}\\ 
\midrule
BLE Positioning Error (m) &5.01 &2.94 &3.13 &4.68  &4.11 &3.97\\

BLE and IMU Positioning Error (m) &4.31 &3.43 &2.57 &4.71 &3.21 &3.65\\
BLE Room Level Accuracy (\%) &88.1 &91.4 &92.7 &89.2  &90.2 &90.3\\
BLE and IMU Room Level Accuracy(\%) &90.7 &90.6 &93 & 90.4  &91.6 &91.2\\
\bottomrule
\end{tabular}
}
\end{table}
\fi

% Link to the figrues: 

%https://docs.google.com/presentation/d/1hVZNv_-l1oIPU7xFDO6U97Er62Li5K9vr0CvcT3AcWI/edit#slide=id.p

\autoref{tab:PositioningErrorOverWindowSizes} displays the positioning errors (in meters) and room-level localization accuracies (in \%) at each area.
Overall, when using BLE and IMU fusion, the proposed method (\autoref{tab:PositioningErrorOverWindowSizes}, 6th and 8th rows) showed 3.9 m positioning error and 87.1\% room-level localization accuracy.
This was a significant improvement for both absolute position error (1.56m) and room-level accuracy (4.9\%) compared to the baseline method (\autoref{tab:PositioningErrorOverWindowSizes}, 2nd and 4th rows).

\section{Discussion}

% \subsection{Indoor Localization Performance}
\textit{Indoor Localization Performance:}
For the proposed method (\autoref{tab:PositioningErrorOverWindowSizes}, 5th and 6th rows), when only using the BLE-based approach, the average localization error across the entire study site was 4.22 m.
When BLE and IMU were fused for localization, the localization error was reduced to 3.9 m, showing that dead reckoning information can compensate for errors in BLE-based localization coming from fluctuating RSSI and unequal coverage of signals across space.
The room-level localization (\autoref{tab:PositioningErrorOverWindowSizes}, 7th and 8th rows) was very effective, with an average accuracy of 87.1\%  when fusing BLE and IMU, which improved 5.7\% from when only using BLE only.
We consider the fine trajectory available from IMU helps to localize subjects near the borders between each area, as the dead-reckoning method can compensate for fluctuating RSSI when the subject is not moving.

\autoref{fig:Qualitative} shows qualitative results while following one subject in our study site. 
Left figure (\autoref{fig:Qualitative}, a) shows the comparison between localization results with the baseline method (\autoref{fig:Qualitative}, a, \textcolor{red}{Red}) and the proposed method (\autoref{fig:Qualitative}, a, \textcolor{purple}{Purple}) against ground-truth locations (\autoref{fig:Qualitative}, a, \textcolor{blue}{Blue}).
Localization results from the baseline method are severely affected by missing detections as it only considers instantaneous BLE beacon hits for trilateration (\textcolor{red}{Red} box).
The proposed method could track the subject well within 4 meters as shown in \autoref{tab:PositioningErrorOverWindowSizes}.
Right figure (\autoref{fig:Qualitative}, b) shows the comparison between the localization results from the proposed method from BLE-only (\autoref{fig:Qualitative}, b, \textcolor{red}{Red}) and BLE-IMU fusion (\autoref{fig:Qualitative}, b, \textcolor{green}{Green}).
Concerning tracking a subject while moving, IMU fusion demonstrated slightly better performance (30 cm) with BLE-only localization results, as shown in the quantitative results.

\begin{figure}
    \centering
    \begin{adjustbox}{width=1.\columnwidth,center}
    \begin{tabular}{c c}
        \includegraphics[align=c, width=0.8\linewidth]{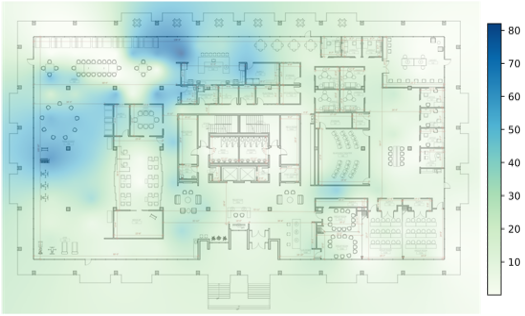} & 
        \includegraphics[align=c, width=0.8\linewidth]{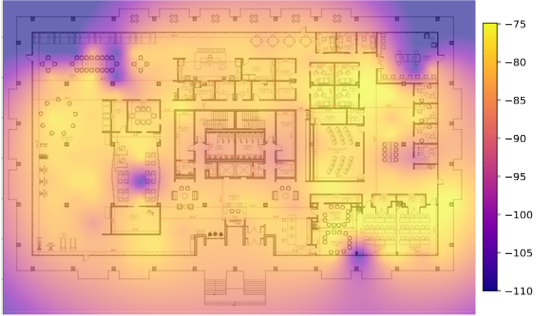} \\
        (a) & (b)
    \end{tabular}
    \end{adjustbox}
    \vspace*{-0.1in}
    \caption{Coverage heatmaps of RSSI signals across study site.
    The (a) number of hits and (b) average RSSI signal observed in study site depends on the density of edge devices and their surrounding structures.}
    \vspace*{-0.0in}
    \label{fig:CoverageEP6}
\end{figure}

\iffalse
\begin{figure}[t]
\centering
\hspace*{-1cm}\includegraphics[scale=0.39]{figures/EP6_coverage.pdf}
\caption{The experiment involved holding a BLE beacon and moving to 131 different locations within the facility, spending 30 seconds at each location. We calculated the following metrics: (a) the number of hits received by the beacon at each location, (b) the number of hits received from unique Raspberry Pis, (c) the average RSSI value for each location, and (d) the number of unique RPis that had frequent hits (i.e., more than 1 hit in 2 seconds). Subsequently, the acquired data underwent interpolation to generate heatmaps based on the measurements obtained from the 131 evaluated locations.}
\label{fig:CoverageEP6}
\centering
\end{figure}
\fi

% \subsection{Impact of Edge Device Distribution}
\textit{Impact of Edge Device Distribution:}
\autoref{fig:CoverageEP6} shows the heatmap representations of the signal reception across the study area, which is collected at the equally spaced 131 locations in our study site.
The signal reception and strength across our study site are inconsistent, with some regions having weak or unstable connections, such as Right Corridor and Lounge areas where high positioning error was showing \autoref{tab:PositioningErrorOverWindowSizes}.
RSSI can fluctuate due to environmental factors, such as walls or furniture, which can block or redirect the signal~\cite{hajiakhondi2020bluetooth}. 
As expected, local regions with more edge devices, such as Left Corridor, Kitchen, and Activity Area, have more number of hits (\autoref{fig:CoverageEP6}, a) and higher average RSSI (\autoref{fig:CoverageEP6}, b), which means more robust and more consistent signal strength from beacons.
As a result, those areas demonstrated lower positioning errors in \autoref{tab:PositioningErrorOverWindowSizes} compared to other regions.
Signal coverage analysis explains the failure of standard trilateration in our study site, which depends on the accurate distance measures between each edge device and beacon.

\section{Conclusion}

This work proposes an open-source and scalable indoor localization approach fusing BLE and IMU wearable sensors using distributed edge computing systems in large indoor spaces. 
Our analysis shows that, especially for large indoor space (over 1600 m$^2$) with complex structures, unequal distribution of edge devices can result in inconsistent coverage of signals that can significantly challenge the BLE-based localization approach.
Yet, the proposed adaptive trilateration approach that uses the temporal density of hits from BLE beacons can robustly localize the position of multiple people with an average error of 4 m across the entire study space, also showing 87\%  accuracy for room-level localizations.
Integrating IMU-based dead-reckoning with BLE-based localization further enhanced the system's accuracy. 
In our future work, we expect to deploy the proposed work to analyze space navigation behaviors for individuals with MCI, which is known to be an indicator of cognitive impairment ~\cite{ghosh2022machine} .

% This work proposes an open-source and scalable indoor localization approach fusing BLE and IMU wearable sensors using distributed edge computing systems in large indoor spaces. 
% Our analysis shows that, especially for large indoor space (1600 m$^2$) in our study, unequal distribution of edge devices can result in inconsistent coverage of RSSI signals that can significantly challenge BLE-based localization approach.
% Yet, the proposed adaptive trilateration approach that uses the temporal density of hits from BLE beacons can robustly localize the position of multiple people with average error of 4 meters across entire study space also showing 87\% accuracy for room-level localizations.
% The integration of accelerometer data using complementary filters further enhanced the accuracy of the system. 
% In our future work, we expect to deploy the proposed work to analyze space navigation behaviors for individuals with MCI which is known to be an indicator for cognitive impairment.~\cite{ghosh2022machine} 

% These findings have significant implications for the application of BLE 4.0 and IMU-based localization systems in medical facilities and home settings, particularly for monitoring patients with mild cognitive impairment.
% While precise room interior location could not be achieved, our results confirm that accurate room-level localization is possible with an average location error of 3.65 m and room-level accuracy of 92.7\%.

% \input{acknowledge/main}

\printbibliography

\end{document}